\documentclass[11pt]{article}

\usepackage{setspace}

\usepackage{tikz} 
\usepackage{hyperref}
\usepackage{longtable}
\usepackage{lscape}
\usepackage{amsmath}
\usepackage{bmpsize}
\usepackage{graphicx}
\usepackage{qtree}
\usepackage{xytree}
\usepackage{array}
\usepackage{mdframed}
\usepackage{apacite}

\title{Comparative Opinion Mining:  a Review}

\author{Kasturi Dewi Varathan\thanks{Author to whom correspondence should be sent. Full address: Department of Information System, Faculty of Computer Science \& Information Technology, University of Malaya, 50603 Kuala Lumpur, Malaysia. Email:  kasturi@um.edu.my. Telephone: +603-7967 7022 ext: 2512. Fax: +603-7957 9249.} \\ Department of Information System \\ University of Malaya \\ Kuala Lumpur \\ Malaysia \and Anastasia Giachanou, Fabio Crestani \\ Faculty of Informatics \\ Universit\`a della Svizzera Italiana (USI) \\ Lugano \\ Switzerland}

\begin{document}
\date{\today}

\maketitle

\newpage
\doublespacing

\begin{abstract}

Opinion mining refers to the use of natural language processing, text analysis and computational linguistics to identify and extract subjective information in textual material.  Opinion mining, also known as sentiment analysis, has received a lot of attention in recent times, as it provides a number of tools to analyse the public opinion on a number of different topics. Comparative opinion mining is a subfield of opinion mining that deals with identifying and extracting information that is expressed in a comparative form (e.g.~"paper X is better than the Y"). Comparative opinion mining plays a very important role when ones tries to evaluate something, as it provides a reference point for the comparison. This paper provides a review of the area of comparative opinion mining. It is the first review that cover specifically this topic as all previous reviews dealt mostly with general opinion mining.  This survey covers comparative opinion mining from two different angles. One from perspective of techniques and the other from perspective of comparative opinion elements. It also incorporates preprocessing tools as well as dataset that were used by the past researchers that can be useful to the future researchers in the field of comparative opinion mining.


\end{abstract}

\newpage

\singlespacing


\doublespacing

\section{Introduction}

Social media has changed the way people communicate~\cite{Qualman10}. Many people are exchanging opinions, emotions, ideas, and critics on social media platforms, as they are freer to express how they feel about a particular aspect or entity. Besides that, studies reveal that they spend a significant amount of time on these media~\cite{Buzetto12}.  This is proven by a vast amount of opinions being posted about movies, musics, and views on certain aspects, people or events on blogs and forums or review sites~\cite{Romero11}. The decision people make is very much influenced by the opinions and emotions posted on these media~\cite{Bar11}.  Brightlocal's local online review survey reveals that 88\% of customers trust online reviews to make their decisions\footnote{\url{https://www.brightlocal.com/learn/local-consumer-review-survey/}}.  The findings reveal the significance of reviews and every 9 out of 10 people read online reviews. This made sentiment analysis and opinion mining an important field of research in every domain of the present time~\cite{Liu12}.

The use of these opinions has become a crucial aspect that needs to be tackled~\cite{Asur10}. Besides expressing his or her own positive or negative opinions about an entity or aspect, a person can also express opinions through comparing similar entities or aspects. Take the sentence \textsf{''}picture quality of iPhone 6 is better than Samsung Galaxy Note 3\textsf{''}. In this paper, we would like to use these kind of comparative opinions. In the past, information on competitors were gathered from formal media such as newspapers, yearly reports, surveys made in order to gauge the market perspective, etc. Nowadays, business organizations are relying on real time customers' reviews in making strategic decisions. This kind of comparative opinion mining contributes a lot to competitive intelligence in helping business organization in identifying risk and potential market at early stages. This in a way helps organizations in making strategic decisions to enhance their sustainability and growth. Besides that, it also helps organization to build online reputation and helps individuals in the decision making process. 

A review on comparative opinion mining is very much needed as there is an exponential growth of people who rely on online reviews for their decisions. This paper will be able to reveal the techniques used so far and to what extend the research in this area managed to progress. It also helps to identify the strengths and weaknesses of each methods that were explored so far, based on the elements (sentence, entity, relation and feature) of comparative opinion mining. This survey is also catered for future researchers to focus on some of the research gaps that exist in each of the elements of comparative opinion mining.

Thus, this paper focuses on comparative opinion mining in which the first section of the paper will give a brief introduction on opinion mining and comparative opinion mining. The main content of this paper focuses on comparative opinion mining techniques and mining comparative elements which includes mining comparative sentences, entity detection, relation detection and feature detection. Resources available for comparative opinion mining are also discussed in order to help the future researchers to be aware and utilize these resources. This survey covers research on comparative opinion mining for the past 9 years (2006-2015). We have covered around 38 papers in this survey that focused on comparative opinions. The survey focuses not only on just English, like most of the other surveys.  It also focuses on other languages especially Chinese and Korean language. The papers of other languages that was published in English venues are included in this survey. However, we will only concentrate on research papers that focus on comparative opinions obtained from customers reviews, blogs or forums.  Studies made on comparative statements in general text such as web documents, news or any other scientific text are not within the scope of this study because they may contain facts rather than opinions~\cite{Bos6,Park12,Yang11,Huang8,Sun6,Zhang12,Fiszman7,Huang11,Yang11,Yang9}. Finally, we present the different performance measures that are frequently used in the literature of comparative opinion mining. 

\section{Opinion Mining}

In this section, we present an introduction and a brief description for the tasks of opinion and sentiment analysis and comparative opinion mining. It is meant to show the substantial differences between the two areas of research. The section concludes with a description of how this review is organized. 

\subsection{Opinion and Sentiment Analysis}

Social media platforms (e.g., Facebook, Twitter, Google+, several blogs) enable people to express and share their thoughts and opinions on the web in a very simple way. User generated data are important as they contain valuable information about the public view on any topic. However, looking through the vast amount of data to extract useful information is nearly impossible. To this end, researchers have started developing approaches that can automatically mine and analyze opinionated information within a huge amount of data.  

\textit{Opinion mining}, also known as \textit{sentiment analysis}, is one of the emerging fields focused on developing methods that can automatically detect opinionated information and determine the polarity of the opinion towards a specific target. The opinion target is usually a named entity such as an organization, individual or event. More formally, Liu and Zhang~\cite{Liu12} define an opinion as "a subjective statement, view, attitude, emotion, or appraisal about an entity or an aspect of an entity from an opinion holder". The same definition states that "an entity is a concrete or abstract object such as product, person, event, organization which can be represented as a hierarchy of components, sub-components, and their attributes". For example, in the sentence "The photos’ quality of my new Nikon camera is excellent" the entity for which an opinion is expressed is the Nikon camera, the aspect of the entity is the photos’ quality and the polarity of the opinion is positive.

Opinion miming is a valuable tool for a variety of applications that require understanding public opinion. One typical example is business intelligence. This refers to enterprises that can use opinion mining to capture the views of customers about their services or products and use this information to improve their quality~\cite{Inui08}. In addition, potential customers of a product can consult past experiences of other users to decide whether buying the product or not~\cite{Bickart01,Morinaga02}. Another application, known as government intelligence, refers to using opinion mining by government to understand what citizens really need and want and act promptly~\cite{Arunachalam03}. Finally, opinion mining can be used in politics for understanding what voters think~\cite{Laver03}. 

Opinion mining has been studied on many media including reviews, forum discussions, blogs and microblogs. There are numerous articles that have been published for opinion mining and sentiment analysis. Two long and detailed surveys on opinion mining were presented some time ago by Pang and Lee~\cite{Pang08} and Liu and Zhang~\cite{Liu12}. 

\subsection{Comparative Opinion Mining}

Comparison plays an important role to evaluate something, it could be in terms of product, services, people, action, etc. In order to achieve the objective of being able to compare objects especially from web 2.0 technologies that incorporate user generated data, we first need to identify and classify which are the statements that belong to the comparative and the non comparative categories.  Just mining opinion and sentiments alone is insufficient, because this will only show how much people talk and how they feel about certain products or services. This leads to incorrect judgment because these reviewers are highly likely not to have any experience on other products or services. On the other hand, comparative opinion is mainly provided by people who have experience related to the products or services they are comparing and it certainly reveals information which are more precise on the similarities and differences between products or services. These kind of evaluations are very important in gauging customers' views because they manage to reveal precise rather than vague information that could be otherwise retrieved from purely direct opinion based reviews. Many people analyse reviews in order to help them make decision.  For example,
{\small
\begin{verbatim}
Example 1: "Samsung Note 3 is good in terms of performance"
Example 2: "Samsung Note 3 is good in terms of performance compared 
            to iPhone"
\end{verbatim}
}

From these 2 sentences, it clearly shows that Example 2 will give more precise information to the future purchaser for helping him to choose one product compared to another. This is the kind of statement that need to be retrieved and analysed. Comparative opinion mining has also become a niche area of opinion mining since we are overloaded with huge amount of user generated data.  Comparative opinion statements comprise of only 10\% from the total opinionated text~\cite{Kessler13} and these comparative statements provide precise information to the users. Thus, much information could be derived in which business organizations and individuals are expected to be the key players who will benefit from comparative opinion mining.

\subsection{Organization of this Review}

This review focuses on comparative opinion mining. It is divided into two main parts which are comparative opinion mining techniques and comparative opinion mining elements. This is like looking at the two sides of the same coin. Section~\ref{techniques} talks about comparative opinion mining techniques exploring research from the perspective of the techniques. On the other hand, Section~\ref{elements} reviews research works from the perspective of the opinion elements (sentence, entity, relation, feature).  There research work based on each of the elements is covered in detail.  Section~\ref{resources} instead covers the resources that are used and available to be used by the future researchers for researching in this area, including available datasets and tools. Conclusions and future work are given in Section~\ref{conclusions}.

\section{Comparative Opinion Mining Techniques}
\label{techniques}


Different techniques have been employed to address the task of comparative opinion mining. These techniques can be roughly divided into the following classes:

\begin{itemize}

\item{Machine Learning}
\item{Rule Mining}
\item{Natural Language Processing}
\end{itemize}

A Machine Learning approach employs a machine learning method and a set of different features to address different tasks of comparative opinion mining. The machine learning methods can be further divided into supervised and unsupervised learning methods. A Rule Mining approach considers methods that are based on association rules and sequential patterns. The rules can be either manually defined or automatically mined from text. The Natural Language Processing (NLP) approaches include methods that apply semantic or syntactic analysis of the text. 

In the rest of this section, we briefly describe the methods that have been applied to address the different tasks of comparative opinion mining. This section is organized based on the approach used regardless of the task being tackled.  

\subsection{Machine Learning Approaches}

Machine Learning is a subfield of Computer Science that studies methods and algorithms that can learn from data and make predictions~\cite{Mohri12}. In comparative opinion mining, approaches from the field of machine learning are usually employed to address different tasks. For example, classifiers are developed and trained with the aim to distinguish comparative sentences from non-comparative sentences. Support Vector Machines (SVM), Na\"{i}ve Bayes and Conditional Random Fields (CRFs) are some of the most popular approaches for comparative opinion mining. 

\subsubsection{Supervised Learning}

Supervised learning includes methods that learn a function given a set of training data. The goal of those methods known as classifiers is to predict the class attribute for the unlabeled data. The classifiers are trained based on a set of features extracted from the training data. Two of the most popular features are the opinion terms (love, great, hate etc.) and the Part-Of-Speech tags (POS) that are used to annotate the words with their syntactic behavior (i.e. verbs, adjectives, adverbs).

One of the most popular text classifier is the {\em Support Vector Machines (SVM)}. According to this method, training data are marked as belonging to one category and mapped into the space so the categories are as wide as possible. SVM has been used in many applications including mining comparative sentences. Wang et al.~\cite{Wang15} employed SVM to build a model that could classify sentences as comparative or non-comparative. Keywords, sequence patterns, and manual rules were used as features to train the model. Their approach was applied to Chinese comparative customer reviews. 

Liu et al.~\cite{Liu13b} also employed SVM to identify Chinese comparative sentences. The SVM classifier was trained using class sequential rules, comparative words and statistical feature words. SVM was also used by Li et al.~\cite{Li11} to classify the polarity of the review sentences. Tkachenko and Lauw~\cite{Tkachenko14} used SVM for comparative sentence and aspect detection. 

A multi-class SVM was proposed by Xu et al.~\cite{Xu9} for identifying and categorizing the comparative relations. In contrast to the typical SVM classifiers that are only applied for two classes, the multi-class SVM can be applied for more than two classes. Xu et al.~\cite{Xu9} proposed to train a classifier for every category and then to assign the classifier with the higher score as the prediction category. Linguistic features including entity types, words and POS tags were used as features to train the model.  

The {\em Conditional Random Fields (CRFs)} proposed by Lafferty et al.~\cite{Lafferty01} are a type of statistical modeling method. In contrast to a typical classifier, CRFs also consider the labeling context to identify relationships between data. CRFs were used by Xu et al.~\cite{Xu11} to extract comparative relations from customer reviews. More specifically, they proposed a graphical model based on a two-level CRF to extract and visualize the comparative relations between products. Complicated dependencies between relations, entities and words, and the interdependencies among relations were used as features. Their algorithm was applied on Amazon customer reviews, epinions.com, blogs, SMS and emails. One interesting conclusion from this study was the usefulness of the comparative relation map for enterprise risk management and decision making. 

CRFs was also applied by Liu et al.~\cite{Liu13a} for extracting comparative elements from the sentences that were already identified as comparative. In their study, they applied CRFs to identify the product name and the target attribute in the comparative sentences. Words, POS, distance from the comparative word and domain-specific words were used as features for the element extraction. Another work that applied CRF was presented by Feldman et al.~\cite{Feldman7}. Feldman et al. employed CRFs to detect terms that frequently appeared with a specific product. For this reason, they built one CRF model for POS tagging and one for chunking. The proposed CRF model was trained on the CoNLL-2000~\cite{Tjong00} corpus. 

Another classifier that is frequently used by researchers in comparative opinion mining is Na\"{i}ve Bayes. {\em Na\"{i}ve Bayes} classifier is one of the most popular classifiers used for text classification~\cite{Mohri12}. The model is based on the Bayesian theorem and performs well when there is a large number of dimensions. In text classification, Na\"{i}ve Bayes follows a bag-of-words approach to compute the posterior probability of a class. In other words, the text is considered as a bag of words without considering grammar and word order. Na\"{i}ve Bayes was employed by Jindal et al.~\cite{Jindal6a} to classify sentences as comparative or not. Jindal et al.~\cite{Jindal6a} used class sequential rules as features. Na\"{i}ve Bayes classifier was also used by Tkachenko and Lauw~\cite{Tkachenko14} to detect comparative sentences and entities of interest. 


\subsubsection{Unsupervised Learning}

Unsupervised learning is a type of machine learning algorithm used to detect hidden structure in unlabeled data. {\em Clustering} is one of the most popular unsupervised approaches. Clustering aims to group a set of data in such a way that similar data will be in the same group. {\em K-means} clustering is one of the most widely known clustering methods. K-means aims to group the observations into k clusters in which each observation belongs to the cluster with the nearest mean. 

In the literature of comparative opinion mining, unsupervised learning is under-explored. To the best of our knowledge, applying unsupervised learning on comparative opinion mining has only been considered by Tkachenko and Lauw~\cite{Tkachenko14} who proposed a generative model for comparative sentences. Their proposed model called CompareGem was able to jointly model comparative directions at the sentence level and ranking at the entity level. Gibbs sampling~\cite{Casella92}, a Markov chain Monte Carlo (MCMC) method, was used for hidden variables inference. CompareGem was evaluated on both supervised and unsupervised settings for relation and entity detection. K-means was used as a baseline in the unsupervised setting. 

%
%


\subsection{Rule Mining Approaches}

{\em Association rule mining} is a popular data mining method for discovering interesting relations between data. When applied to text mining, its main objective is to discover strong rules and patterns between terms. {\em Sequence pattern mining} is another type of rule mining. In contrast to association rule mining, it considers the order of terms within a sentence. According to the definition given by Agrawal et al.~\cite{Agrawal93} the rules in association rule mining are defined as $X \Rightarrow Y$ where $X,Y$ are subsets of items.

This mining model is general and can be applied in many domains. For example, it is used in market analysis for a more profitable product promotion, in medical diagnosis to assist physicians to cure patients and in bioinformatics for analyzing DNA sequences. Association and sequential rules have also been extensively applied in Web and text documents.

Associations were considered by Ganapathibhotla and Liu~\cite{Ganapathibhotla8} with the aim to identify the preferred entity in the comparative sentences. Pros and Cons reviews were used to determine the sentiment orientation of the comparative word and the feature. Comparative words, compared features, compared entities, and a comparison type were used as features in their method. They also proposed one-side association (OSA) measure to compute the association between a comparative word and a feature. 

{\em Class Sequential Rules (CSR)} is a type of sequential pattern mining in which each data sequence is labeled with a class. CSR was proposed by Jindal and Liu~\cite{Jindal6a} to discover frequent patterns in comparative sentences. The sequence patterns, which were used as features, were based on POS tags and comparative key phrases. The sentiment of each pattern was then determined by the sentiment of the manually identified key phrases. Their experiments on news articles, consumer reviews and forum discussions showed that combining CSR with manual rules could effectively identify comparative sentences. 

A CSR-based approach was also considered by Liu et al.~\cite{Liu13a} who proposed rule-based and CSR-based methods to identify comparative sentences. Comparative words, adverbs and syntactic patterns were used by the rule-based approach. For the CSR-based approach they combined sequential rules mining and machine learning techniques. The identified comparative sentences were then classified into three classes according to their sentiment orientation. CSR was also considered by Liu et al.~\cite{Liu13b} who used those rules to train the classifier to detect comparative sentences written in Chinese. The results were significantly better when CSR features were considered during the training. 

Rule mining was also used by Liu et al.~\cite{Liu5} for developing a framework with the aim to analyze and compare consumer opinions. Liu et al.~\cite{Liu5} proposed a system called Opinion Observer that could be used for comparing opinions on different products. Supervised rule mining was applied to extract product features that were part of the comparison. The generation of rules was based on POS tags and n-grams. Experiments on a dataset extracted from Epinions showed that the proposed system was effective.  

In addition to CSR, Jindal and Liu~\cite{Jindal6b} also applied {\em Label Sequential Rules (LSR)} to identify comparative sentences. LSR are defined as $ X \Rightarrow Y $ where $Y$ and $X$ are sequences and $X$ is produced by $Y$ by using wildcards to replace some of the items of $Y$. Jindal and Liu~\cite{Jindal6b} employed POS and comparative keywords as features. Experiments on consumer reviews, forum postings and news articles showed that LSR managed to effectively identify the comparative sentences. 

Instead of mining rules from text, a number of researchers manually defined a number of rules that they considered being indicative of comparison. This approach was applied by Gu and Yoo~\cite{Gu10} who proposed using sentence pattern rules and comparison words to detect comparative sentences that were written in Korean. To this end, they defined sentence rules indicative of comparative sentences. In their approach, they used comparison words and POS tagging. They evaluated their method on comments about restaurants. 

A pattern-based approach was used by He et al.~\cite{He12} on reviews about mobile phones written in Chinese. He et al. defined six patterns to extract the comparative relations from reviews. A comparative and a superlative feature lexicon were used to detect the polarity of the features. The results showed that the pattern-based method managed to significantly improve the baseline method which was on the label sequence rules.  

A similar approach was proposed by Kurashima et al.~\cite{Kurashima8} who used language patterns to detect entities and extract comparative relations. In their study, they manually prepared language patterns to extract entities and opinion. Experimental results on reviews about movies showed that the proposed method can effectively extract comparative relations. 

\subsection{Natural Language Processing Approaches}
Natural Language Processing (NLP) is a field of computer science that focuses on developing methods that can process and analyze written or spoken language. NLP methods can be applied to analyze language at two different levels: syntactic analysis and semantic analysis. Syntactic analysis parses and analyzes the syntax of sentences. Semantic analysis aims to identify and analyze the meaning of words, phrases and sentences. Here, we present the studies that applied a syntactic or a semantic analysis to the comparative opinion mining task. 

\subsubsection{Syntactic Analysis}

Dependency parsing is a type of syntactic analysis and aims to identify the syntactic relations between words. The output of dependency parsing is the syntax tree that reflects the syntactic structure of a sentence. Figure~\ref{fig:syntaxTree1} shows an example of a syntax tree of the sentence "Good diagrams are an important tool". 

\begin{figure}
\centering
{\LARGE Insert Figure 1 here}
\caption{The parse tree of the sentence "Good diagrams are an important tool"}
\label{fig:syntaxTree1}
\end{figure}

Dependency parsing has been used in many domains including computational linguistics, information retrieval, opinion mining, etc. Different information can be derived by the dependency parsing including dependency grammar graph. Dependency grammar graph identifies and assigns grammatical roles to the words of a sentence. For example, it is used to identify the subject, the direct object, the indirect object etc. To make this more clear we provide the dependency tree of the sentence "Nick bought a camera." in Figure~\ref{fig:dependency} which shows the syntactical relations within the sentence.

\begin{figure}
\centering
{\LARGE Insert Figure 2 here}
\caption{The dependency tree of the sentence "Nick bought a camera."}
\label{fig:dependency}
\end{figure}

Identifying the grammatical roles is very useful in comparative opinion mining as it can recognize the directions of comparative relations or mine feature opinion pairs. To this end, Sun et al.~\cite{Sun9} proposed an automated system based on dependency grammar graph and evolution tree. Their system could compare and recommend products to the customers from both subjective and objective perspectives. Dependency grammar graphs were used to mine feature opinion pairs. Then, the authors proposed an evolution tree to visualize how a specific product evolved. The evolution tree was also proposed to recommend customers the products about which customers had a good opinion. 

In addition to dependency grammar graph, Xu et al.~\cite{Xu11} extracted also syntactic paths in order to extract comparative relations. Syntactic paths were useful in detecting comparative relations because they were usually expressed with similar syntactic patterns. Their experiments on Amazon customer reviews showed that syntactic features are useful for extracting comparative relations. Dependency trees were also employed by Liu et al.~\cite{Liu13b} who analyzed $2,000$ comparative sentences with the aim to identify similar syntactic patters. After this analysis, they concluded that a similar syntactic structure is followed in the two parts of the comparative sentences. 

\subsubsection{Semantic Analysis} 

Semantic analysis tries to address the problem of language ambiguity by extracting the meaning of the sentence. One popular method for semantic analysis is {\em Semantic Role Labeling (SRL)}. SRL aims to detect the semantic roles and relationships associated with the verbs or the predicate of a sentence and the semantic arguments. In SRL, the predicate represents an event and the arguments represent the participants of this event. Typical examples of semantic roles are Agent, Recipient, Patient etc. For example, in the sentence "The teacher talked to the student in the front desk", "the teacher" is the Agent, "talked" represents the Predicate and "the student in the front desk" is the Recipient. Assigning semantic roles is of high importance for several domains such as Question Answering and Summarization. 

Kessler and Kuhn~\cite{Kessler13} proposed a model based on SRL to detect the entities and the predicate in comparative sentences. Using a standard pipeline from SRL, they identified the predicate and the arguments. The arguments were then classified in terms of being positive or negative arguments. A regularized linear logistic regression classifiers was used for the classification. The proposed approach managed to outperform the baselines on a dataset containing blog posts about cameras and cars~\cite{Kessler10}.

Hou and Li~\cite{Hou8} used the SRL approach to extract comparative relations in Chinese text. They used SRL to detect six elements in comparative sentences: holder, entity1, comparative predicates, entity2, attributes and sentiments. POS, phrase type, position to predicate, the comparative predicate were used as features in the SRL. As a last step, they trained a CRF classifiers using those features to label the comparative sentences. Their experiments on two different datasets of $200$ comparative sentences each, showed that using SRL can effectively mine Chinese comparative sentences. 

{\em Semantic network analysis} is a semantic approach that is based on a network to reflect the semantic relations between concepts. Figure~\ref{fig:semantic} shows an example of a semantic network on which a set of entities and their attributes are depicted. The most well known example of semantic network is WordNet~\cite{Fellbaum98} which is a large lexical resource of English. Words are grouped together into sets of synonyms which are linked to each other by semantic and lexical relations. WordNet has been also used in comparative opinion mining to detect synonyms and extend the list of comparative words~\cite{Jindal6a}, to find relations between features and opinion words~\cite{Kim9} and identify synonyms of features~\cite{Liu5}. 

Another example of semantic network is SenticNet\footnote{\url{http://sentic.net/}}, a knowledge base that includes a set of semantics, sentics, and polarity associated with $30,000$ natural language concepts. SenticNet has been used to address concept-level sentiment analysis by leveraging semantics and linguistics~\cite{Cambria2015}. Poria et al.~\cite{Poria2014} used SenticNet for aspect extraction from product reviews. In their work, they used common-sense knowledge and dependency trees to extract implicit and explicit aspects, a task that is very useful in comparative opinion mining.

\begin{figure}
  \centering
  {\LARGE Insert Figure 3 here}
      \caption{Example of a semantic network}
\label{fig:semantic}
\end{figure}

Semantic network analysis was employed by Kurashima et al.~\cite{Kurashima8} who proposed a graph-based method in which the nodes represent the entities and the edges reflect the relations between two entities. According to the authors, the graph structure was appropriate because it could model the behavior of a potential customer who goes around until he finds the best entity. The authors also proposed the graph centrality score to measure the importance of an entity compared to the remaining entities. 

Kim and Zhai~\cite{Kim9} also proposed a semantic approach to address the contrastive opinion summarization problem. This task was focused on summarizing opinions by detecting sentences with contrastive perspectives on the same product. For example, the reviews "The screen quality of MacBook Air is very good" and "MacBook Air comes with a poor screen quality" are contrastive. Kim and Zhai~\cite{Kim9} addressed the contrastive opinion summarization problem with an optimization framework. The proposed method was based on several similarity measures. They also proposed two additional measures for this task, representativeness and contrastiveness. Representativeness measures how well the summary represented the original text whereas contrastiveness reflected the similarity between positive and negative sentences. The authors used reviews from Amazon to evaluate their method. 

Li et al.~\cite{Li11} employed unified graphical model to represent each aspect of a relation. Each graphical model is based on all the comparative reviews obtained for a particular aspect. Their model was evaluated on reviews about phones and audio players. Finally, Fujimoto~\cite{Fujimoto12} investigated mathematically potency-magnitude relations to model the degree of difference between two objects. To this end, they proposed a Q-magnitude Relation Map (Q-Map) and a Priority Message-Class Map (P-Map). Those models used evaluation target size and evaluation scale size to reflect how the relations change. One interesting conclusion of this study was that non-equal gradable messages that included intensified comparisons were most likely to change. 

\subsection{Conclusions}
After presenting the techniques that have been used in the different tasks of comparative opinion retrieval, we observe that most of the research employed a rule based or machine learning approach. Pattern-based approach is suitable for comparative opinion mining since those sentences follow a specific pattern. However, they have to be combined with other methods to obtain a better performance. Unsupervised or NLP approaches which are still under-explored in this field could be suitable. For example, including synonyms in order to improve the performance of the approaches would be an interesting direction. 

\section{Mining Comparative Opinion Elements}
\label{elements}

This section explains comparative opinion mining from a different point of view compared to the previous section. In this section, comparative opinion mining is reviewed from the perspective of elements.  In comparative opinion mining, research was seen in comparative opinion sentence detection, entity detection, relation detection and feature detection.  The advantage of this categorization is that it identifies the attributes that we have focused and helps in narrowing down the scope of the research works. It also helps the future researchers to identify the research gaps easily and more clearly. Finally, we would like to highlight that the works done on each of these elements aim to attain different objectives and therefore their comparison is not easy. Thus, this section is focused on exploring each one of these elements together with the techniques used. In the past, linguists spent some significant number of years in studying comparatives constructs in English language. Some had defined the problem as a universal quantifier over degrees of gradables~\cite{Lerner92}, whereas others defined it as ordering of objects with some degrees of gradables~\cite{Kennedy5}. There is research which focused mainly on defining the syntax and semantics of comparative  constructs from a linguistic perspective~\cite{Moltmann87}.  Just by analysing comparative opinion sentences with comparative words is ineffective. For example, \textsf{''}In order to be the best, we have to work hard\textsf{''} is not a comparative opinion statement although it contains the comparative adverb \textsf{''}best\textsf{''}. There are also comparative sentences that will not contain any of the comparable words in lexicons, but belong to comparative opinion statements.  There is also research on comparative sentiment identification through the use of logics~\cite{Von84, Ballard88, Rayner90}. Work reported in~\cite{Friedman89} used comparatives in question answering systems. Subsequently,~\cite{Staab97} used a multi-layered interpretation model which combines syntactic and semantic approaches. All these studies covers the logics of comparative statements and not on detecting comparative opinions. The following subsections will explore the elements of comparative opinion mining, which are comparative opinion sentence detection, entity detection, relation detection and feature detection.

\subsection{Comparative Opinion Sentence Detection}

In this section, we look at past research on mining comparative opinion at sentence level. The first work on computational methods in comparative sentence extraction in opinionated text was seen in~\cite{Jindal6b}. However, research by Zhai et al.~\cite{Zhai4} with cross-collection mixture model for comparative text mining exist even before~\cite{Jindal6b}, but their research focused on how to retrieve different common themes across all the available text collections and not on sentence level identification. On the other hand, Liu et al.~\cite{Liu5} studied analysing and comparing competing products based on opinions rather than on comparative sentences as a whole. Meanwhile, Kim et al.~\cite{Kim9} worked on summarizing contradictory opinion from opinionated text and not on comparative opinion sentences.  For the past 9 years (2006-2015), the focus of the research on comparative opinion sentence detection covers machine learning, associative rule and lexicon approaches.  In total, there are 6 research papers on these approaches. The research focuses on supervised machine learning technique~\cite{Wang15} and rule mining~\cite{Liu13b}. The associative rule approach in comparative sentence identification was initiated by Jindal and Liu and also used by Liu et al.~\cite{Jindal6a, Jindal6b, Liu13a}.  Meanwhile Wei et al.~\cite{Wei14} explored a lexicon based approach. The details on each one of these approaches are covered in depth in the following subsections. 

\subsubsection{Associative Rule Mining Techniques}

All research that falls under the category of associative rule mining uses {\em Class Sequential Rule (CSR)}. CSR was used by Jindal and Liu and Liu et al. in distinguishing the comparative and the non comparative text. Jindal and Liu~\cite{Jindal6a} have grouped comparatives into 4 types (non-equal gradable, equative, superlative and non-gradable).  "Canon is better than Nikon" is an example of non equal gradable type which contains expressions that shows greater or lesser than.  Meanwhile equative type shows the equality between the entity or entities. An example would be: "Canon is as good as Nikon". For superlative comparative type, the relation of one entity is compared to all the others, as an example: "Canon is the cheapest of all". On the other hand, non gradable types do not grade the objects they are compared with. Comparative sentence such as "Some of Canon's features are different than Nikon's" is an example of non gradable type. Jindal and Liu~\cite{Jindal6a} managed to distinguish between comparative and non-comparative sentences by using all the different types of comparative sentences based on a set of 83 keywords. They carried out several experiments based on keywords, Na\"{i}ve Bayes and SVM. The performance of Na\"{i}ve Bayes on both CSR and manual rules show a significant improvement in precision but as expected, recall dropped.

Jindal and Liu~\cite{Jindal6b} focused on gradable comparative sentences which comprises of non-equal, equative and superlative, and neglecting the non-gradable sentences. Non gradable sentences comprises of sentences that do not have explicit grading. For example, "Hotel A has air condition, but Hotel B does not have". This kind of comparison is implicit and cannot be graded. They also extended their own studies which worked on CSR and took advantage of Label Sequential Rule (LSR).  They filtered sentences that are not comparative by using the same set of keywords that they used previously using supervised learning approach. Both techniques, CSR and LSR, performed reasonably well.

Liu et al.~\cite{Liu13a} also used CSR and rule based method on Chinese comparative sentences.  The rules are compiled based on syntactic patterns and words comprising of comparative words, comparative content words and relative degree adverbs.  On the other hand, the CSR method integrates class sequential rule mining and machine learning techniques. A list of comparative words is created and these words are used to generate the sequential dataset from the training corpus. A multi minimum support was used to extract all the patterns that match the minimum threshold. This leads to an increase in their precision, but dropped recall. The reasons for low recall is due to the incomplete list of comparative keywords used, the complexity of the text, wrong spelling and punctuations. 

All the works in the area of associative rule mining used sets of keywords and syntactic patterns in identifying comparative sentences. This only tackles the problem at the basic level since lexicon do not portray the meaning of these sentences. A sentence may have keywords and syntactic pattern that belongs to comparative category, but may not necessarily be a comparative opinion sentence. Thus, a semantic method is needed to handle this problem and improve the performance of the classifications obtained so far.

\subsubsection{Machine Learning Techniques}
Using machine learning approaches to detect comparative sentences has emerged in recent years.  This can be seen with the study conducted by Wang et al. on supervised approaches that uses {\em SVM\/} and Liu et al. on rule mining techniques. Wang et al.~\cite{Wang15} worked on comparative sentence identification in Chinese text.  They divided comparative sentences into two types, which are comparative sentences comprising comparative keywords and non comparative keywords. The keyword search technique is used in order to identify sentences with comparative indicators and syntactic technique was used to identify comparative sentences that do not have comparative keywords on Chinese Opinion Analysis Evaluation (COAE2012). A keyword lexicon was created that consist of 102 words and 30 syntactic sequences. SVM was used for filtering non comparative sentences that contain keywords. In order to handle superlative comparative sentences, 45 patterns were also constructed. The precision performance obtained is the best precision in comparative sentence detection so far. 

A supervised rule mining technique was used by Liu et al.\cite{Liu13b} in identifying comparative sentences in Chinese Opinion Analysis Evaluation (COAE2013) and online product reviews sites. They constructed three kinds of sentence structure based on these datasets and the characteristics of comparative sentences. There is a slight similarity between this research and the previous that used SVM~\cite{Wang15}, in which both used syntactic structure for differentiating the sentences.  The difference between these two works is the use of the sentence structures. Liu et al.~\cite{Liu13b} used sentence structure for identifying comparative and non comparative sentences. On the other hand, Wang et al. used to identify comparative sentences that do not have comparative keywords.  The study also reported that high recall was achieved in identifying explicit comparative sentences compared to implicit ones.  To handle the implicit comparative sentences, researchers mined more information by using dependency relations and as a result, they found two parts of the comparative sentences that have similar syntactic structures.  A statistical technique was also used to compute the similarity dependency relation and to exclude all the sentences that were below a determined threshold. Their approach managed to give a satisfactory recall and not on satisfactory precision. This is due to most of the sentences managed to fit into the broad template that was created.  As a consequence, they proposed a SVM classifier that included a {\em Class Sequential Rule}, comparative words and statistical feature words. They also extended their experiment on SVM by incorporating sentence structure and similarity of dependency relation. The result of this experiment showed an increase of all the performance measures compared to their own previous experiments. 

From the above, it showed that supervised learning is used for identifying comparative sentences in Chinese text only. To the best of our knowledge, there is no research in comparative sentence detection which uses supervised techniques on English text or that covers other languages. On the other hand, strangely there is no research on unsupervised learning of comparative sentence identification. Future research is welcomed to explore machine learning approaches in comparative opinion sentence identification.

\subsubsection{Lexicon/NLP Techniques}

There are two papers that use syntactic techniques in comparative sentence detections in Chinese text~\cite{Gu10, Wei14}. Gu et al.\cite{Gu10} used sentence structures and statistics in identifying comparative sentences. The sentence that contains graded comparative words such as 'than', 'more', 'far' and 'more' were extracted and classified as comparative sentences

Meanwhile, Wei et al.~\cite{Wei14} proposed a method that uses some usual conceptual arrangements of words in a sentence and the structural characteristics of comparative sentences. However, there are few drawbacks in both of these studies: they did not use word segmentation, and this leads to identifying non comparative sentences as comparative.  Besides that, they have failed to handle negation as well as comparative sentences that had omitted "than" but used "is".  

Research on lexicon based approach should evolve. Semantic based technique would be a good area that needs to be explored as no research was found on this technique so far. 

\subsubsection{Conclusions}

The research on mining comparative sentences is still at its infancy. Most of the techniques used are based on some kind of {\em pattern based\/} approach or {\em machine learning\/} approach with regards to the languages covered (English, Chinese). Surprisingly there are only two research papers on English text~\cite{Jindal6a, Jindal6b} while the rest are on Chinese text~\cite{Gu10, Liu13a,Wang15,Liu13b,Wei14}. Future research on comparative sentences should venture into unsupervised approaches and lexicon based semantic approaches. Besides that, the collections of comparative keywords need to be increased in order to achieve better recall and precision since all past research uses sets of keywords for identifying comparative sentences. One easy approach could be to include synonyms for keywords as well as incorporating domain knowledge. But again, using just words in identifying comparative sentences is a conventional approach. Research should flourish beyond this aspect.

On the other hand, there is a big research gap on comparative opinion mining. Past research only focuses on mining comparative sentence and not on mining comparative opinion sentences. For example, "Everest is the highest mountain in the world" is a comparative sentence which contains fact rather than opinion. On the other hand, "my car is faster than yours" is an opinionated sentence. Unfortunately, none of the research managed to differentiate between comparative facts and comparative opinions. All the papers covered so far are giving a general methodology that could be used for differentiating comparative with the non comparative text. When researchers detected comparative sentences, they basically managed to identify comparative sentences that exist in their dataset (it could be reviews, blogs or forums that belongs to opinionated text and non opinionated text like news). Jindal and Liu~\cite{Jindal6a, Jindal6b} and Liu et al.~\cite{Liu13b} uses reviews as well as news dataset for evaluating their works. In comparative opinion retrieval, this could be biased as they have considered news dataset that are quite prone to have factual data rather than opinionated text. Besides that, assuming all reviews are opinionated is also wrong. A review may contain facts as well as opinions. For example, "KLCC is taller than the parliament building of Malaysia"(fact), "Samsung is more trendy than iPhone"(opinion). Much research on comparative sentence mining manually annotated their sentences into comparative and non comparative. But none has explicitly mentioned that they annotated it based on only comparative opinion sentences~\cite{Jindal6a, Jindal6b, Liu13a, Wang15}. Wang et al.\cite{Wang15} have eliminated direct opinion sentences and only concentrated in comparative opinions. However, they did not omit comparative sentences that do not contain opinions. On the other hand, Ganapathibhotla \& Jindal~\cite{Ganapathibhotla8} worked on opinion in comparative sentences in which they made use of opinionated comparative words, context and application domain in detecting opinion in comparative sentences. However, their focused more on identifying preferred entities rather than comparative opinion sentences. As a consequence, they did not perform any evaluation to support their findings on mining opinion in comparative sentences.  In conclusion, research on mining comparative opinion sentences is still at its infancy.

\subsection{Entity Detection}

Entity detection is one of the main components of comparative opinion mining as it provides valuable information to consumers on what are the other options that are available for the products that they are looking.  It might also help business organizations in order to know who are their current rivals. Besides that, the overall structure of the market can be identified and viewed.  The entity can be represented as product names, service names, person names, etc. Entity detection basically covers identifying brand names or model names. 

In this section, we focus on how entities are identified in comparative opinion sentences. A closer look is needed since we are not handling general text with standard names or properly mentioned abbreviations or proper language text, with correct spellings and grammar. It is not right to expect reviews given by consumers to not have all these. Dealing with informal language requires more specific techniques compared to formal text.

\subsubsection{Machine Learning Techniques}

Research on {\em Conditional Random Fields\/} (CRF) is implemented in detecting entities in comparative sentences. Liu et al.~\cite{Liu13a} worked on retrieving two entities in a single comparative statement. For a comparative statement that contains a single entity, a "NULL" value is replaced. An example of such statement would be "the picture quality is good as that of Canon". They classified comparative sentence to contain four elements which are subject, subject attribute, object and object attribute. An example of such sentence is "Mercedes’s speed is faster than BMW’s speed". In this case, the subject is Mercedes, subject attribute is speed, object is BMW and object attribute is speed. They proposed CRF to identify product names in Chinese opinion text.  The product names are often mentioned at the end of a comparative sentence. Thus, CRF is used to identify entities by using only the end word in each of the element phrase as the head word which contains the entity name that exist in the domain knowledge. The performance of classification that uses head word shows better precision and recall compared to baseline approach that tries to classify entities by using each words in a sentence. Very low recall is achieved and this is caused mainly by the limited number of words in the list. Besides that, this research also showed that higher domain knowledge delivers better results. In terms of machine learning, only CRF is used in retrieving entities in Chinese text. CRF technique could also be used by researchers who deal with other languages to see whether this technique can be applied to non Chinese text.  Other techniques from machine learning still remain unexplored in entity detection from comparative opinion text.

\subsubsection{Lexicon Techniques}

This section discusses all the relevant literature on lexicon based approaches in detecting entities in comparative sentences. {\em Lexicon based approaches\/} are more popular compared to other approaches. This could be seen from the number of papers published on this approach.  Xu et al.~\cite{Xu11} worked on mobile phone data sets and had also used their own lexicon dictionary that contains mobile phone names and attribute to detect entities in a sentence. Since mobile phones may have different abbreviations for the same product, for example, "iPhone 6 Plus" is also known as "6+" or "iPhone6+" in reviews. Thus, they have also included all the possible abbreviations in building their lexicon. Anaphora which address the issues of word referring back to a word used earlier such as “it”, “they”, “she”, “he”, etc. were also tackled by adopting “closest-first” method.. However, they have realized the importance of performing post processes steps that caters for products which are mentioned differently but referring to the same entity. For example, "Samsung S2" could be referred to as "S2". This kind of different entity names needs to be normalized in order to avoid duplication of entities. Since the focus of their research is not on entity normalization, they have neglected this phase.

Tkachenko \& Lauw~\cite{Tkachenko14} used a data driven approach in detecting entities in comparative sentences. Their study revealed that in order to identify comparable entities, these entities must have been compared at least once in the opinionated text. Besides that, there exist entities which are compared indirectly with one and another. To accommodate this, they found that if entity1 is compared with entity2 and entity2 is compared with entity3 then entity1 can be compared with entity3. They have scoped their research to deal with comparative sentences that contains 2 entities only. Since there is no Name Entity Relation (NER) for the entities that they were detecting, they created their own dictionary based on product titles and employed a dictionary matching approach.

Feldman et al.~\cite{Feldman7}, identified an entity by emphasizing on the entities that are mentioned together in a sentence. Named entity recognition methods were used to accomplish this task. A graph was generated in order to identify which entity co-exist with one another. Overall visualization on how the market is structured based on entities from the perspective of consumers can be obtained. They have also used a dictionary based approach which collects all the terms used to refer to the same entity.  However, this approach is not sufficient to accommodate to comparative opinion text since it contains many variations and it is not possible to make sure that all these variations are covered by the dictionary. They realized that brand names are easier to be identified, compared to model names and model numbers. This is because model number could be written in many formats or variations such as only numbers are written in the review without stating the model name. Another issue that needs to be taken into serious consideration is that names could be misspelled. To tackle all these problems, researchers devised a process of entity detection into 2 stages. First stage was matching the sentences with list of search strings from the web and use regular expressions in identifying the entities.   Once these entities are identified, the second stage filtered out and modify the wrongly extracted entities from the list. In another word, it removed falsely recognized entity. All comparative sentences' sequence of tokens, POS tags and NP chunks that satisfied minimum support value were retrieved. From these, frequent sequence sets were identified. These patterns have been filtered to accommodate only patterns that contain specific words or part of speech. Identifying patterns with parsing failed to give fruitful results due to grammatical or spelling errors in the opinionated text. Their findings also show that simple pattern matching gives better result compared to parsing.

Kessler et al.~\cite{Kessler13} proposed {\em Semantic Role Labelling\/} (SRL) for detecting entities. They had used predicate with three arguments compared to the normal two. These three are entity which is evaluated positively, entity which is evaluated negatively and the aspect in which these entities are evaluated in a comparative sentence. All these three arguments were filled as semantic roles. The dataset was annotated by human on positive entity, negative entity and aspect. It was then used as a training set for the SRL system. They were aware that identifying entity does not focus on comparing against a set of comparative keywords that lies between the entities alone. Thus, their research used Part of Speech (POS) in entity identification. This approach filled up the gap that exists in achieving better recall as it managed to identify comparative words which are not stated in the set of comparative keywords as what had been used by other researchers such as~\cite{Jindal6b}. As a consequence, any comparative opinion text could adopt this method without much hassle. The difference between this research and the other research is that entities are associated with roles and not with predicates. Since they worked with comparative sentences that contain 2 entities, they successfully detected both entities. Their results were compared with~\cite{Jindal6b} that used rule based approach which will be discussed in depth in the next section and their findings reveal that they outperformed all performance measurements (recall, precision and F-score).  

Research on entity detection based on lexicon approaches managed to cover a number of techniques compared to other approaches. Techniques on dictionary based, data driven, regular expressions and semantic role labeling were used in detecting entity in comparative opinion. 

\subsubsection{Rule Base Techniques}

The rule based technique is also used in entity detection. Jindal and Liu~\cite{Jindal6b} worked with two entities and used {\em Label Sequential Rules\/} (LSR) which uses noun and pronoun in extracting entities in comparative sentences.  These entities were extracted with a set of 83 comparative keywords composed of 4 POS tags(Comparative Adjective such as "taller", "bigger", Comparative Adverb such as "faster", "more frequently", Superlative Adjective such as "best", "most desirable"  , Superlative Adverb such as "earliest", "hardest" ) and $79$ keywords referring to indicative words such as outperform, beat, etc.  The rules were used to extract the components of the relation that consist of entities and features. LSR reports good result with very high precision. However, the dataset that they used for testing contained only $564$ gradable comparatives which included three types of comparatives (non-equal gradable, equative and superlative).   Although the dataset contained mostly reviews and forum discussions, it also contained random news articles that may include factual information rather than opinionated information. Thus, the performance results obtained are questionable, due to the biases that exist in the dataset.

\subsubsection{Text Mining Techniques}

He et al.~\cite{He12} worked with a mobile dataset similar to~\cite{Xu11} using lexicon based approach. They used a text mining approach in identifying entities in comparative sentences. In this research they categorized entities in comparative sentences in a different way compared to the previous approaches. Each of the comparative sentences contains a subject and an object and this subject and object can be from mobile brands or mobile models. The Jaccard coefficient was used as a mining technique in identifying the entities in comparative sentences. The similarity value obtained from this technique was used in differentiating the subject as well as the object in comparative sentence. Entity with high similarity value in comparative sentence is categorized as a subject and entity with lower value as object.

\subsubsection{Conclusions}

Not much research was found that solely catered for entity detection in comparative opinion mining.  Research dealt with just utilizing the existing approach to entity retrieval, since the focus of the present researchers lies in other area of comparative opinion mining. The accuracy of entity retrieval is important since it is one of the main backbone of comparative opinion mining. It is pointless to find the relation between entities or finding the features of entities when the entities are wrongly identified. Extracting entity in comparative opinion text requires more precise techniques compared to general text. It is tedious to cater for some fixed rules or fixed number of lexicons. It is also tedious to involve human efforts to accomplish good results in entity detection as the data we are dealing with is huge and requires automated mining.  Data variations become a more complex problem to handle since opinionated text contains many variations in writing styles and is ambiguous. The use of dictionary, POS tags analysis or domain knowledge alone will not be sufficient to solve the problem of identifying entities in a comparative opinion. Up to now, only CRF techniques of machine learning are used for identifying entities in comparative text. There are many other techniques that could be explored such as Support Vector Machine, Hidden Markov Model, etc.  Other text mining approaches could also be used for identifying entities. Research done so far is narrowed to a maximum of two entities in a comparative opinion sentences. -	Research done so far is narrowed to a maximum of two entities in a comparative opinion sentences. For example, "Rolex is expensive compared to Swatch". This sentence contains only two entities, which are Rolex and Swatch.  Future research should tackle comparative sentences that contain more than two entities.  Besides that, the problem of pronoun resolution needs to looked into more carefully as it contributes significantly to entity detection.

\subsection{Relation Detection}

Relation detection is an important aspect as it helps in knowing the relation that exist between entities and how to rank these entities. This kind of information is very valuable for business organizations in knowing the strengths and weaknesses of their products compared to their competitors. Besides that, individuals will also be able to benefit from relation detection since they will be able to know what are the options available and what suits them the most based on their requirements. In this section, we look at past research that handled relation detection and to what extent this relation detection had been explored in comparative opinion mining.

\subsubsection{Machine Learning Techniques}

This section explores research that used machine learning techniques in relation detection. Tkachenko et al.~\cite{Tkachenko14} had proposed a generative model called CompareGem (COMParative RElation GEnarative Model) that tackles both sentence level as well as entity level comparisons using both supervised and unsupervised approach. It is a model that managed to fill up the gap that exist in the pipeline approach. For pipeline approach, it managed to show the entity that is superior than the other. However, there are some limitation, for instance, in Figure~\ref{fig:Pipeline Example} it shows a pipeline example.

\begin{figure}
  \centering
  {\LARGE Insert Figure 4 here}
      \caption{Pipeline example}
\label{fig:Pipeline Example}
\end{figure}

In this example, we cannot conclude whether E is superior to D or the other way around. Besides that, they realized that using bag of words alone is insufficient to achieve this goal. Thus, this research successfully achieved it through a sigmoid ranking function.  Sentence level comparison gave an output of which entity is superior to the other.  Significant results were obtained for CompareGem at sentence level. Meanwhile, entity level comparison determines the overall ranking of these two entities that were compared in the comparative sentence. An entity that is preferred in a comparative statement does not mean it would be preferred when compared to the overall comparative statements. This kind of comparison is similar to the one reported in~\cite{Kurashima8} which used graphical model in detecting relations but that take into consideration the overall comparative relation in finding the superior entity by using query. For further details on~\cite{Kurashima8}, refer to graphical model subsection. Although CompareGem managed to give a different perspective on the overall relation between comparative sentences, it failed to rank the top most entities.  In their research, they did not cater for the equality that exist between entities because they claimed that it would not produce any effect. For example, A is better than B, B is better than C and D is equal to A. The probability that D is better than B is higher.  However, in order to measure which entity is superior, equality plays an important role.  Meanwhile, although CompareGem managed to produce highest accuracy for entity ranking, the results obtained are quite similar to the baselines. However, their supervised configuration managed to produce better result compared to their unsupervised approach.

\subsubsection{Lexicon Techniques}

Lexicon based techniques are used in detecting relation in comparative opinion mining. There are four research works that used this technique in their relation detection. Jindal et al.~\cite{Jindal6b} is the first research that proposed mining comparative relation in opinionated text. They defined comparative relation to be represented as follows:-
{\small
\begin{verbatim}
(relation word, features, entity 1, entity 2, type). 
\end{verbatim}
}
The relation word is identified with a set of comparative keywords, entity 1 and 2 are the respective entities that exist in the comparative statement and type represents one of the three different types that had been identified: non-equal gradable, equative and superlative. Non-equal gradable expresses the preferences, for instance, "worse than" or "better than". Equative comparative relation treats both entities as equal and superlative sentences identifies relation between entities in which one of the entity is preferred compared to all others. Although these types of comparative relation classification cover most of the comparative sentences, it failed to handle all types of comparative sentences that belong to non-gradable. Non-gradable comparative relation focuses on comparing more than two entities without any explicit grading. In this study, they only handle comparative statements with only one relation. This caused drawbacks since users opinion may contain more than one relation that could exist between entities or features. Besides that,~\cite{Jindal6b} only focuses on categorizing comparative sentences into the three different categories that were mentioned earlier. They also failed to make use of the direction of the relations which identifies the preferred entities. On top of that, since they used Label Sequential Rule, their rules are insufficient to be adopted by other domains.

Hou \& Feng~\cite{Hou8} represent the structure of comparative relation as attributes, entity 1, operator and entity 2.   An example of attributes would be price, speed, distance, capacity, etc. Entity 1 and entity 2 comprises of products, services, persons or locations. Meanwhile operator is one of {>,<,=, ˄,˅} in which ">" represents that entity 1 is better than entity 2 and "<" represents vice versa. The "=" operator shows that both entity 1 and entity 2 are equal. Entity 1 is the best is shown using "˄" operator and "˅" means that entity 1 is the worst.  In this study, they have used a huge number of comparative sentences:1800 sentences and 277 different comparative predicates. These sentences are used in training the CRF model. However, complex sentences that did not manage to go through parser are excluded in the evaluation process. This would have led to some important sentences missing in the evaluation process and thus producing higher precision since the complex sentences are discarded automatically.  Meanwhile, they only catered for sentence level relation detection and not on overall relation. This means, an entity can be considered as superior, inferior, equal, worst or best in a comparative sentence and not by considering the overall comparative sentences.

Semantic Role Labelling (SRL) was used by Gu et al.~\cite{Gu10} in Korean text. It is the first research that studies relation extraction in Korean text. For the very first time, their research managed to identify the comparative directions in a relation and also identified the superior entity that exist in comparative statements. The structure of comparative Korean text consist of:
{\small
\begin{verbatim}
(subject, target, feature and superiority)
\end{verbatim}
}
\noindent In this comparative sentence, "Fabric A is more expensive than fabric C", the subject is "fabric A", "fabric C" is the target, feature is the price, superiority is "fabric A". Predicate information that lies next to the superiority or inferiority comparison word is utilized to indicate the relation that exists in comparative sentences.  Thus, they performed a rule based approach to identify which is the superior entity. The rules are applied to all predicates which are located on the right hand side of a comparative word.
The dataset used for this study is unique as they used only comments on restaurants.  Besides that, their evaluation centered only on retrieving comparative sentence based on four different comparative keywords which are "than", "compared to", "far" and "more". The usage of this very limited amount of keywords shows that  the research needs to be enhanced further in terms of keywords as well the rules used. They have basically tackled only the general part of comparative relation in opinionated text, although they achieved surprisingly good result in terms of precision. 

He et al.~\cite{He12} build a comparative relation lexicon which contains comparative words. They identify relation through comparative words used in a sentence. This study suggested six matching patterns to extract comparative words from opinionated text. The matching pattern identified are purely on syntactic, very simple and straightforward. Figure~\ref{fig:table} shows the matching patterns. 

\begin{figure}
\begin{center}
%
%
%
%
{\LARGE Insert Figure 5 here}
\end{center}
\caption{Matching patterns} \label{fig:table}

\end{figure}

These patterns are too general to be adopted in detecting relation in comparative opinion sentences.  Besides that, they are also insufficient to cater for the complexity and the variations of Chinese language. It also clearly shows that no attention was given to comparative opinions, but only on the part of speech of comparative sentences. 

\subsubsection{Graphical Models Techniques}

{\em Graphical models\/} in comparative sentences based on opinion are first used by Kurashima et al.~\cite{Kurashima8}. They created a model that represents the relations that exist only between entities without including their features. The entities were also ranked against their competitors.  Their approach is based on query, whereby the users can make a query on single entity or multiple entities, the results will be based on overall comparative relations. This approach is somewhat similar to the one proposed in~\cite{Tkachenko14}. In this research, they managed to find all the competitors (entities) and aspects which were similar to the searched query. The searched query basically provided an idea on which entity cluster the user was interested in. A graph was generated based on the comparative relation in which nodes represented entities and edges represented the relations. All the nodes relating to the queried entity or entities were retrieved. Their result shows the ranking of all the competitors based on the users' query. Thus, the customers could have an idea of the available options and of the best option that they could pick.

Xu et al.~\cite{Xu9} proposed a SVM based map for comparative relation. They were the first one to introduce comparative relation map. This map consists of entities, the attributes and the relation between each one of these attributes with other entities. Their model managed to identify if a comparative relation exists and  differentiates it into three sentiments categories which are "better", "worse" and "same". They broke the norm by dealing with three entities in which handling these amounts of entities is a multi class classification problem. However, the relations extractions that they handled were quite straightforward. Besides that, the dataset that they obtained from Amazon contained only 217 comparative relations which were quite low to train their model which was based on a machine learning approach. They should have used a larger dataset for training which could have given a better performance in their research work. 

Xu et al.~\cite{Xu11} also used the graphical model to interpret the relation that could exist between entities. This graphical representation managed to show all the entities that could exist in the dataset. This kind of relation visualization helps in supporting decisions made by business organizations in tracing the information about their competitors' strengths and the strengths of their own products or services. They identified how to differentiate comparative relation from other existing relation extraction models. Their findings revealed that comparative relation belonged to higher order relation in which it contained four entities or arguments. On the other hand, the existing relation contained only two entities. Besides that, comparative relation contained direction in which such directions do not exist in normal opinion statements. For example, "I like iPhone" is a normal opinion without any directions. On the other hand, "I like iPhone compared to Samsung", is a comparative statement that contains direction.  To accommodate these needs, these researchers proposed a two level CRF model with unfixed interdependencies that is a more powerful modelling tool compared to the existing CRF model.   Although they managed to retrieve relations that were useful for business intelligence, their results are not very promising. 

Graphical models were also used by Li et al.~\cite{Li11} in relation detection. The difference between their model and others is in the comparisons made. They have compared relation based on aspects such as design, feature, performance and ease of use. Each aspect is represented with a unified graphical model that takes into account all the comparative reviews. This unified approach helped the customers to have a clear picture on which product seemed to be the best, based on superiority scores of the aspects. K-Mean of clustering were used to group the comparative sentences in reviews and community based question answering (cQA) pairs. For polarity classification of these sentences, SVM is used. The results obtained higher average score in using both reviews and cQA compared to just reviews.  Although their model gave a different perspective on how comparative relations are performed, not all products could use the same aspects to make comparisons. For example, using performances aspects to compare fabrics, clothes, handbags, book, etc. are not feasible. Besides that, these aspects cannot be expanded to be used in non product domains, such as services, since aspects will be different.

\subsubsection{One Side Association Techniques}

{\em One side association techniques\/} are used in relation detection of comparative opinion mining.  Ganapathibhotla \& Jindal~\cite{Ganapathibhotla8} worked on identifying which is the preferred entity in a comparative opinion mining sentence. This research expected to benefit the consumers in deciding their best options. They emphasized on sentimental aspects of a sentence in order to determine if either Entity1 or Entity 2 is preferred. This sentiment is derived from context dependent opinion, which indicates whether the comparative word gives positive or negative opinions. For example, "more" does not mean to be positive all the time, it depends on which context this word is used. For instance, "Device A has more power consumption than Device B ". In this case, the comparative word "more" and "power consumption"  give negative value to an opinion. Thus, they derived the following rules:

        \begin{equation*}
        \begin{array}{c@{\qquad}c}
        \text{Increasing comparative word +  Positive} 
        \Rightarrow
       \text{Entity 1 preferred} \\ 

		\text{Decreasing comparative word +  Positive}    
                \Rightarrow
                \text{Entity 2 preferred}   \\ 

		\text{Increasing comparative word + Negative} 
                \Rightarrow
                \text{Entity 2 preferred} \\ 

		\text{Decreasing comparative word + Negative} 
                \Rightarrow
                \text{Entity 1 preferred} \\ 
                
        \end{array}
        \end{equation*}

Since, context dependent information were used in this research, they proposed one side association method to identify the number of times comparative words and features co-occur.  The base forms of these words are also included. Besides that, synonyms and antonyms, which were retrieved from WordNet, were also used. The count increased every-time comparative word and feature (including the synonyms) co-occured in positive comparative statements. The count was also increased if antonyms of comparative word and feature together with its synonym co-occur in negative statements.  Besides features, adjective or adverbs were also used to detect the preferred entity. On the other hand, these researchers tackled negation statements as well. They had compiled a list of 26 negation words and negated the rules mentioned earlier in detecting the preferred entity. There are times in which negation in comparative opinion statements will not show any preference to any of the entities. An example of this kind of statement would be "Printer A's performance is not better than that of Printer B".  The result obtained showed significant improvement compared to the baselines.

\subsubsection{Conclusions}

There are 9 papers that worked on relation detection in comparative opinion mining in this past 9 years (2006-2015). From these 9 research, 2 research were on machine learning approaches, 2 are from rule based approaches and the others used of lexicon based approaches. None of the research was focused on unsupervised approach. Exploring unsupervised approaches for an instance clustering technique could group entities that belong to the same product or services category. From this, the user will be able to identify the relation that could exist between the products or services. There are also many other techniques that have yet to be ventured on, such as other text mining approaches that could be explored to identify which entities are always being compared with. Business organizations could benefit from this kind of association since it could identify the top most rivals that exist or allows consumers to know what are the best options available for a product or services that they are looking for.  Besides these approaches, the languages that were covered in comparative opinion mining so far are only on English, Chinese and Korean text. Future work may also make use of some past approaches to focus on other languages.

\subsection{Feature Detection}

Features information on comparative opinion mining is very important. In fact, it is often useful to look at products' or services' features.  Detailed information on certain entities could only be found in these features. For example, "I like Obama more than George Bush", failed to indicate based on which attribute (feature) the "liking" is. This kind of statement do not provide much information to the readers, especially in decision making purposes. On the other hand, for instance, "Obama's administration capability is better than George Bush's" clearly shows that  the preference is based on Obama's administration capability.  The value of an entity is appreciated based on the feature in which the entity is compared rather than just the entity itself.  In this section, we will focus on how features are detected and explored in comparative opinion mining.

\subsubsection{Statistical Techniques}

This section focuses on {\em statistical approaches\/} used in feature detection and how they are implemented.  Sun et al.~\cite{Sun9} handled features in comparative opinion text by building a sentiment based product features database. The comparison between products was made based on the structured data that were extracted and saved in the database. Visualization based on tree models was used to identify options as well as criteria such as lowest price or best selling items.  Moreover, customers would prefer to have some detailed comparison which includes the feature aspect of the products that they wish to compare rather than some general information.  To accommodate this, they proposed a recommender system that took into account all the previous reviews and suggested products based on feature comparisons. For instance, a customer could post a query such as "Suggest me a car that is better than BMW". Their system would list all the cars which are better than BMW based on each of the common features between BMW and other cars.  Sentiment values on features based on past reviews were also included in suggesting products. Rather than just giving binary values sentiments, they used SentiWordnet to weigh each of the sentiment strength accordingly. This definitely gives a positive impact to the comparative opinion text they are handling since stronger sentiment opinion has a higher value than the weaker one. 

Their system was tested using reviews obtained from Amazon with 18 different products' features.  Furthermore, the evolutionary tree that was proposed managed to give a clear picture of the product evolution process and served as a guide to customers for getting the latest or better products in the market. These researchers looked into the utilization of features in comparative opinion mining from totally a different angle. A sample of their evolutionary tree is shown as the Figure~\ref{fig:syntaxTree}.

\begin{figure}
  \centering
  {\LARGE Insert Figure 6 here}
  \caption{Evolutionary Tree}
\label{fig:syntaxTree}
\end{figure}

On the other hand, they failed to handle synonyms of features which were mentioned differently in the review.  For example, "photo", "image", "picture" were all representing the same feature. Although this looked simple, their performance would have been much better with this integration.

\subsubsection{Pattern Matching Techniques}

{\em Pattern matching\/} techniques were explored in feature detection. Tkachenko et al.~\cite{Tkachenko14} proposed a generative model that includes information that can be derived on sequence of the features.  These features can be categorized to syntactic feature that analyzes the placement of a word, whether it is before, between or after the mentioned entity and negation features which exist to a targeted word. Since they are focusing on a dataset of digital cameras, four different aspects that were frequently used to compare camera such as functionality, form factor, image quality and price were included as features that needed to be analyzed. Their analysis revealed some interesting points indicated in Figure~\ref{fig:features}.

\begin{figure}
  \centering
  {\LARGE Insert Figure 7 here}
  \caption{Top 5 Most Desciminative Features}
\label{fig:features}
\end{figure}

\noindent The sequence of existence of words is deemed very important. For example, column 1, row 1 of the element of the table (\#1 from \#2) shows that the sequence (\#1 from \#2) is the top discriminative feature in terms of functionality that contains in a comparative sentence that favored entity 1. Another example would be column 8, row 1(\#1 more \#2) reveals that if entity 1 is more than entity 2, then entity 2 is favored. The opposite of this can be seen in column 7, row 1(\#1 less \#2) which favors entity 1. However, since the features are manually identified in the camera domain, there is a possibility for this research to exclude some important features such as size.

\subsubsection{Conclusions}

So far, only pattern matching and statistical approaches were used for feature detection in comparative opinion sentences.  Statistical approaches managed to give the detailed aspects of features and pattern matching enable to identify the preferred entity based on pre identified features.  Research works were seen in using the features as relations and not much research was founded on solely feature detection studies.  There are a lot of aspects that can be looked into in feature detection for comparative opinion mining.  Explicit and implicit feature detection in comparative opinion sentences could be a new research area. Besides that, topic modelling could also be used for identifying features from comparative opinion reviews.

\subsection{Conclusions}

Research on categorizing comparative and non comparative opinion sentences is not being explored yet. It is an important area that needs to be looked into.  This is because comparative opinion sentences are unique by itself. A comparative sentence that contains opinion word need not necessarily be a comparative opinion sentence. Besides that, research on handling non gradable comparative sentences need to be explored as no research is found. Moreover, entity normalization is an important area that needs to be looked into.  Past research used manual insertion by listing the possible abbreviations.  This approach tends to miss some abbreviations which may affect the overall entity identification results.  We found that most research on comparative opinion mining works only on two entities.  Future research should cater for more than two entities.  Many research works are found in relation mining compared to other elements mining.   Although relation seems to be popular compared to the others, there is not enough research done on evaluation metrics in ranking the entities in comparative reviews.  On top of that, feature detection needs to be researched further as it contains the least amount of research works. When these are achieved, more solid and fruitful results could be obtained. 

\section{Performance Measures}
\label{measures}
Comparative opinion mining can be considered as a classification problem since it aims at classifying if a sentence, entity, relation or feature is retrieved or identified correctly.  For this reason, the most frequent measures used in the literature are \textit{accuracy}, \textit{precision}, \textit{recall} and \textit{F-score} that are adopted from traditional classification field.

To better explain these measures we provide one example. Consider we want to measure the performance of a classifier on its ability to detect comparative  sentences. Given a sentence, the classifier annotates as positive the sentences that are considered as comparative, otherwise the sentences are annotated as negative. Figure~\ref{fig:confusion} summarizes the results of the classifier on the test data by providing the number of \textit{True Positives} (TP), \textit{True Negatives} (TN), \textit{False Positives} (FP) and \textit{False Negatives} (FN) instances. TP represents the number of sentences predicted as positive and which were indeed positive, whereas FP is the number of sentences incorrectly predicted as positive. TN and FN have a corresponding meaning for the negative class. Based on this matrix, we now explain the most frequent comparative opinion mining measures.

\begin{figure}
\begin{center}
%
%
{\LARGE Insert Figure 8 here}
\end{center}
\caption{Confusion Matrix} \label{fig:confusion}

\end{figure}

\begin{itemize}
\item \textit{Accuracy}: Accuracy measures how often the method being evaluated made the correct prediction. It is calculated as the sum of the true predictions divided by the total number of predictions. That is: 
\[
Accuracy = \frac{TP + TN}{TP + TN + FP + FN}
\]

\item \textit{Precision}: Precision is calculated as the ratio of sentences that were correctly predicted as positive divided by the total number of sentences that were predicted as positive. That is: 
\[
Precision = \frac{TP}{TP + FP}
\]

\item \textit{Recall}: Recall represents the ratio of sentences that were correctly predicted as positive divided by the total number of the positive sentences and is calculated as: 
\[
Recall = \frac{TP}{TP + FN}
\]

\item \textit{F-score}: Calculating only recall and precision is not enough. A combination of the two measures is more suitable to measure the performance of the methods. F-score combines recall and precision and is calculated as:
\[
F\textnormal{-}score = 2 \cdot \frac{precision \cdot recall}{precision + recall}
\]

\end{itemize}

The above mentioned measures are mostly used for performance evaluation in comparative opinion mining research. These kind of  metrics are important for  future researchers in conducting their evaluations correctly. Besides that, using the right performance metrics will enable  research to be benchmarked appropriately against the past research works.

\section{Resources for Comparative Opinion Mining}
\label{resources}

This section reports on different resources, such as popular collections and preprocessing tools that have been used in research on comparative opinion mining. 

\subsection{Test Collections}

A number of different {\em test collections\/} have been used in the literature of comparative opinion mining. In most of the cases, researchers built their own collections by crawling reviews from Amazon, Epinions, forums, etc. However, in most cases, a detailed description of the dataset is missing and researchers do not make collections publicly available. The lack of standard test collections creates considerable difficulties in comparing the effectiveness of different methods and systems. 

In an effort to cover the lack of standard collections, some researchers built collections for comparative opinion mining and made them publicly available. Here, we present these datasets and indicate where they could be accessed:

\begin{itemize}
 \item J\&L dataset\footnote{\url{http://www.cs.uic.edu/liub/FBS/data.tar.gz}}: the J\&L dataset that was created by Jindal and Liu~\cite{Jindal6b} contains about 650 comparative sentences extracted from reviews, blogs and forum discussions. This test collection was built specifically for the task of detecting comparisons. 
 \item JDPA dataset\footnote{\url{http://verbs.colorado.edu/jdpacorpus/}}: the JDPA dataset was built by Kessler et al.~\cite{Kessler10} and contains 506 sentences about cameras and 1100 about cars. The sentences were extracted from blog posts. 
\item Kessler14 dataset\footnote{\url{http://www.ims.uni-stuttgart.de/forschung/ressourcen/korpora/reviewcomparisons/}}: the Kessler and Kuhn~\cite{Kessler14} dataset contains annotations for 1707 sentences extracted from Epinions. The sentences are about cameras and are annotated with comparisons. 
\end{itemize}

\subsection{Preprocessing Tools}

The task of comparative opinion mining requires a {\em preprocessing\/} phase that includes tokenisation, stemming, POS-tagging and named entity detection. A number of preprocessing tools have been used for such a phase. Apart from these tools, a number of tools that contain implementations of popular methods have been used to facilitate the research. Those include NLP toolkits and tools that contain implementations of machine learning methods. Although, those tools are very important, in many cases they are not explicitly mentioned in papers. Here, we present the tools that have been used in the literature of mining comparative opinions: 

\begin{itemize}
  \item Gate\footnote{\url{http://gate.ac.uk/}}: Gate is a freely available toolkit for language processing that is provided by the Natural Language Processing Research Group of the University of Sheffield since 1995. Gate is a widely used toolkit for text mining. It provides a number of features including sentence tokenization, entity recognition, past of speech tagging, semantic tagging etc. Gate was used by Xu et al.~\cite{Xu11} to detect linguistic features from the reviews which were then used in mining comparative opinions. 
  \item Stanford CoreNLP\footnote{\url{http://nlp.stanford.edu/software/corenlp.shtml}}: Stanford CoreNLP is a framework that provides a set of natural language analysis tools including tokenization, part-of-speech tagging, named entity recognition, parsing, and coreference. Stanford CoreNLP was used by Kessler and Kuhn~\cite{Kessler13} for sentence segmentation and tokenization. 
  \item Stanford Parser\footnote{\url{http://nlp.stanford.edu/software/lex-parser.shtml}}: Stanford Parser is a natural language parser used to detect the grammatical structure of the sentences. For example, it can detect which term is the subject or object of a verb. It provides a number of features including producing dependency trees. Xu et al.~\cite{Xu11} employed Stanford Parser to detect the grammatical roles of the various entities in the sentences.
  \item SVM multi-class\footnote{\url{http://svmlight.joachims.org/}}:  SVM multi-class contains an implementation of the SVM method. It includes a number of features including solving classification and regression problems. Xu et al.~\cite{Xu11} used this tool to build a model that can mine comparative opinions. 
  \item CRF toolbox\footnote{\url{http://www.cs.ubc.ca/~murphyk/Software/CRF/crf.html}}: CRF toolbox is a Matlab toolbox that implements the CRF method. Xu et al.~\cite{Xu11} used this tool to build a two-level CRF model for extracting comparative relations. 
\item WEKA\footnote{\url{http://www.cs.waikato.ac.nz/ml/weka/}}: WEKA is a toolbox that contains the implementation of machine learning methods. Tkachenko and Lauw~\cite{Tkachenko14} used WEKA to train their baselines which were based on SVM and Na\"{i}ve Bayes. 
\end{itemize}

\section{Conclusions and Future Work}
\label{conclusions}

To the best of our knowledge, this is the first work that reviews comparative opinion mining. There are some significant findings that were obtained from this review. Research on comparative opinion mining published in English venues so far were explored only in English, Chinese and Korean languages.  This finding opens the door to future researchers to embark into other languages as well. Besides that, this review also show that the approaches used in comparative opinion are skewed to NLP.  Many papers were found using NLP based approaches in achieving their objectives. There are many other techniques which remained unexplored, such as unsupervised approaches or computational approaches that uses some mathematical theories, etc.  

Research in the area of comparative opinion mining should focus more on comparative opinion and not on comparative statements in order to avoid factual information.  In terms of application areas that were covered so far are mainly focused comparative opinions on products, although very few research was seen using restaurant and bus service review datasets.  The focus on other domains are important as well. Research that handles comparative opinion mining on public figures or politicians is not ventured at all. 

Besides that, there is no research found in spam detection on comparative opinion mining. This is a useful research path that needs serious considerations among future researcher since spams pollute reviews and give the wrong perception to the decision makers who are relying on comparative reviews in making their future decisions.  

Past research also shows that there is no proper dataset that is available for the purpose of experimentation and evaluations in comparative opinion mining. Although Kessler's dataset \cite{Kessler14}, provided annotated comparative sentences, they still failed to check whether it is opinion based or just mere comparative sentences. Many researchers created their own dataset from publicly available reviews. However, this kind of self private dataset made the evaluation to be biased and not transparent. It is also hard for the researchers to benchmark their results with others since the dataset used for the evaluation purpose differs from one another.  This dataset creations will help other researchers to utilize and evaluate their research based on a standardized data.  Information scientist will be able to explore some of the research gaps that exist which have been highlighted in this review.  They may contribute in many angles from building a proper dataset to be used for comparative opinion mining to defining an evaluation measures that managed to rank the entities, sentences and features that lie in the reviews.

Moreover, this survey devise another possible direction that the future researchers should embark which is on the use of time feature pertaining to comparative opinion mining.  So far there is no research found on the use of time and this feature is indeed very useful for trend analysis, brand sustainability, preference analysis and change analysis.  This survey also reveals that more than 60\% of researchers who worked on comparative opinion mining are Chinese researchers. This opens the door for sociological studies to find out the reason on why this group of people are more interested in comparative opinion mining compared to the others.

\section*{Acknowledgements}

This research was partially funded by Swiss Secretariat of Education, Research and Innovation (SERI).

\newpage

\singlespacing

\bibliographystyle{apacite}
\bibliography{refs}

\end{document}